\documentclass[referee]{raa}
\usepackage{graphicx,times}
\usepackage{natbib}
\usepackage{longtable}
\usepackage{chngpage}
\usepackage{array}
\input{epsf.sty}                        
\input{psfig.sty}                       

\begin{document}

   \title{A two-step energy injection explanation for the rebrightenings of the multi-band afterglow of GRB 081029
$^*$
\footnotetext{\small $*$ Supported by the National Natural Science Foundation of China.}
}

  \volnopage{Vol.0 (200x) No.0, 000--000}
   \setcounter{page}{1}

   \author{Yong-Bo Yu  \inst{1, 2}    \and  Yong-Feng Huang  \inst{1, 2}      }

\institute{ Department of Astronomy, Nanjing University, Nanjing 210093, China; hyf@nju.edu.cn
  \and
Key Laboratory of Modern Astronomy and Astrophysics (Nanjing University), Ministry of Education, Nanjing 210093, China
 }



\abstract{ The afterglow of GRB 081029 showed unusual behavior, with a significant rebrightening being
observed at optical wavelength at about 3000 s after the burst. One possible explanation is that
the rebrightening is resulted from energy injection. Here, we present a detailed numerical study of
the energy injection process and interpret the X-ray and optical afterglow light curves of GRB 081029.
In our model, we have assumed two periods of energy injection, each with a constant injection power.
One injection starts at $2.8\times10^{3}$ s and lasts for about 2500 s, with a
power of $7.0\times10^{47}$ ${\rm erg}$ ${\rm s^{-1}}$. This energy injection is mainly engaged to
account for the rapid rebrightening at about 3000 s. The other injection starts at
$8.0\times10^{3}$ s and lasts for about 5000 s.  The injection power is
$3.5\times10^{47}$ ${\rm erg}$ ${\rm s^{-1}}$. This energy injection can help to explain
the slight rebrightening at about 10000 s. It is shown that the observed optical afterglow,
especially the marked rebrightening at about 3000 s, can be well reproduced. In X-ray band,
the predicted amplitude of the rebrightening is much shallower, which is also consistent with
the observed X-ray afterglow light curve. It is argued that the two periods of energy injection
can be produced by the falling of clumpy materials onto the central compact object of the burster,
which leads to an enhancement of accretion and gives birth to a strong outflow temporarily.
\keywords{gamma rays: bursts - ISM: jets and outflows - individual: GRB 081029 }
}

    \authorrunning{Y.-B. Yu \& Y.-F. Huang}            
    \titlerunning{Two-step energy injection in GRB 081029}  
    \maketitle

%
\section{Introduction}           
\label{sect:intro}

Gamma-ray bursts (GRBs) are bright flashes of gamma-rays coming from random
directions in the sky at random times (for recent reviews, see Zhang 2007, Gehrels et al. 2009).
The fireball model is very successful
and popular in view of the fact that it can well explain the main features of GRB
afterglows (Rees \& ${\rm M\acute{e}se\acute{a}ros}$ 1994; Piran 1999; Zhang 2007), which are generally believed to
arise from the interaction of
the fireball with the surrounding interstellar medium (ISM) (${\rm M\acute{e}se\acute{a}ros}$ \& Rees 1997a;
Piran 2000; ${\rm M\acute{e}se\acute{a}ros}$ 2002).
After the amazing coincidence of GRB 980425 and SN 1998bw (Galama et al. 1998),
more and more observational facts have been accumalated (for recent review, see Bersier 2012),
indicating that long GRBs are associated
with Type Ic supernovae. Based on these observations, it is believed that long GRBs
should be due to the collapse of massive stars (Woosley 1993; ${\rm Paczy\acute{n}ski}$ 1998; MacFadyen \& Woosley 1999).
At the same time, it is also widely accepted that short GRBs could be connected with the
coalescence of two compact objects (Eichler et al. 1989; Narayan et al. 1992; Gehrels et al. 2005; Nakar 2007).

With the advance of observational techniques, especially after the launch of the Swift satellite,
many unexpected behaviors are observed in GRB afterglows, such as quick or high amplitude
rebrightenings in optical band, and strong or multiple flares at X-ray wavelength (for recent review, see Zhang 2007).
GRB 081029 is one of the interesting events, which has a remarkable rebrightening
in its optical afterglow light curve. Other examples include GRB 060206
(${\rm W\acute{o}zniak}$ et al. 2006) and GRB 970508 (Sokolov et al. 1998).
These rebrightenings are obviously inconsistent with the simple form of
power-law decay as predicted by the standard fireball model with synchrotron emission
coming from the forward shock of ejecta ploughing into an external medium
(Rhoads 1999; Sari et al. 1999). Many different mechanisms
have been proposed to explain the rebrightenings, including the density jump
model (Lazzati et al. 2002; Dai \& Wu 2003; Tam et al. 2005), the energy injection model
(Dai \& Lu 1998; Rees \& ${\rm M\acute{e}se\acute{a}ros}$ 1998; Huang et al. 2006), the two-component
jet model (Huang et al. 2004; Liu et al.
2008), and the microphysics variation mechanism (Kong et al. 2010), etc. However,
in the more detailed numerical simulations, Huang et al. (2007) argued that the density
jump model is not an ideal mechanism to produce the rebrightenings in optical afterglows.
Holland et al. (2012) also ruled out the possibility that the extremely steep
rebrightening of the optical afterglow of GRB 081029 is resulted from the density
structure in the surrounding environment, due to the fact that it was unable to
reproduce the magnitude of the increase in luminosity. Interestingly,
Holland et al. (2011) reproduced some of the X-ray and optical/infrared rebrightenings
reasonably well with the two-component jet model. In this model, the
early afterglow emission is produced by the narrow, fast component while the
wider, slower component dominates the afterglow after about 3000 s. But their
calculations still failed to reproduce the rapid rise as seen in the ${\rm UVOT}$ data.

Energy injection from late and slow shells seems to be a natural interpretation for
the rebrightening of many optical afterglows. Especially, GRB 970508 exhibited a
late-time flare similar to what is expected from colliding shells (Sokolov et al. 1998).
In this study, we will use a two-step energy injection mechanism to explain the
observed unusual X-ray and optical afterglow light curves of GRB 081029.
In our calculation, we only consider the synchrotron emission, which is the dominant radiation mechanism
that takes place in the afterglow stage, although inverse Compton scattering may also play a role in some
cases (Wei \& Lu 2000; Sari \& Esin 2001). 
The outline of our paper is as follows. The observational facts are presented
in Section 2. The two-step energy injection model, including the dynamics and the radiation process,
are described in Section 3. In Section 4, we calculate the overall dynamical evolution
of the outflow numerically, and reproduce the unusual X-ray and optical afterglow light
curves of GRB 081029. It is shown that the observed rebrightening in the optical band
can be well reproduced. Finally, in Section 5, we summarize our results and give a brief
discussion. We use an assumptive cosmology of $H_{0}$ = 71 km
${\rm s^{-1}~Mpc^{-1}}$, $ \Omega_{M} $ = 0.27 and $\Omega_{\Lambda}$ = 0.73
throughout the paper.


\section{DATA}
\label{sect:Obs}

At 01:43:56 UT on 2008 Oct 29, GRB 081029 triggered the Burst Alert Telescope (BAT)
onboard the Swift satellite and was located at coordinates
$RA(J2000)=23^{h}07^{m}06^{s}, Dec(J2000)=-68^{\circ}10^{'}43.4^{''}$
(Cummings et al. 2008). The peak flux of GRB 081029 measured by the BAT in the
15 --- 150 ${\rm keV}$ band was $(2.8\pm1.3) \times10^{-8}$ erg ${\rm cm^{-2}~s^{-1}}$
with the duration of $T_{90} = 280 \pm 50$ s. The spectrum was best fit by a
simple power law with a photon index of $\Gamma = 1.5 \pm 0.2$ (Holland et
al. 2011), and the redshift measured by the VLT/UVES and Gemini-South from several
absorption features in the host galaxy of GRB 081029 was $ z=3.8479 $ ( Nardini et al. 2011).
The luminosity distance between GRB 081029 and the earth is $3.5\times10^{7}$ kpc for a
standard cosmology with $\Omega_{M}$ = 0.27, $\Omega_{\Lambda}$ = 0.73 and
$H_{0}$ = 71 km ${\rm s^{-1}~Mpc^{-1}}$. A number of ground-based telescopes performed
follow-up observations, providing the multi-frequency light curves of the
afterglow. GRB 081029 was an unusual event with an unusual optical light curve among the GRBs.
One of the most remarkable features of this burst is that the optical light curve has a
significant rebrightening at around 3000 s.


\subsection{Optical Afterglow}
The optical afterglow of GRB 081029 was identified by ROTSE located at the
H.E.S.S. site at Mt. Gamsberg, Namibia, 86 s after the burst. The REM telescope
equipped with the ROSS optical spectrograph/imager and the REMIR near-infrared camera
started observing the optical afterglow 154 s after the BAT trigger in the R, J, and H
bands. GROND, mounted at the 2.2m MPI/ESO telescope at La Silla, started observing the
field of GRB 081029 about 8 minutes after the trigger. A steep rise was observed
in all seven available optical and NIR bands. The Swift/UVOT began observing the
afterglow of GRB 081029 at 2689 s after the BAT trigger, and the afterglow was detected in
the $v$, $b$, and white bands, which was consistent with
the reported redshift of z = 3.8479 (Holland et al. 2012). The R band light
curve shows many interesting features, and is very different from the optical afterglow
of a typical GRB. Firstly, the initial light curve decayed in the normal way with the
simple power-law extrapolation, but the afterglow rebrightened significantly and rapidly
at about 3000 s after the trigger, interrupting the smooth early-time temporal evolution,
which cannot  be explained by using the standard afterglow model. The obtained light curve
confirmed the rebrightening from $r^{\prime}\sim18.6$ magnitude to a peak value of
$r^{\prime}\sim17.4$ magnitude, probably implying a sudden release of a large amount of
energy at late times. Secondly, the optical afterglow light curve became a little flat at
around 8000 s. Finally, the afterglow flattened again after about two days, suggesting either
the presence of an underlying dim host galaxy or a further change in the optical decay
index (Nardini et al. 2011).

\subsection{X-ray Afterglow}
Owing to observing constraints (Sakamoto et al. 2008), XRT and UVOT onboard the Swift satellite
started to follow-up GRB 081029 about 45 minutes after the BAT trigger, but X-ray observations
continued for approximately 10 days. The X-ray afterglow light curve shows a shallow initial decay
followed by a rapid decay, but does not show strong evidence for a marked rebrightening as compared
to the optical afterglow, which casts some doubts on the common nature of the optical and X-ray
afterglow emission. However, it should be noted that there is some evidence for flaring between
approximately 2500 s and 5000 s. The X-ray light curve could be described by a broken power law
($f_{\nu}$ = $t^{-\alpha}$) and the best-fitting model has decay indices of
$\alpha_{1}$ = 0.56 $\pm$ 0.03 and $\alpha_{2}$ = 2.56 $\pm$ 0.09, with a break time of
$t_{b}$ = 18230 $\pm$ 346 s (Holland et al. 2011). The Swift/XRT spectrum can be fit by a single power
law function ($f_{\nu}$ = $\nu^{-\beta}$) with an index of $\beta$ = $0.98 \pm 0.08$. There is
no evidence for any evolution in the power law index at X-ray energies (Holland et al. 2012).


\section{MODEL}

In recent years, Eerten et al. (2010) developed a code for the dynamical evolution of GRB afterglows.
Their calculations include some delicate ingredient and are relatively accurate. But their code
is also relatively complicate. Here we will use the simple equations for beamed GRB ejecta developed
by Huang et al (1998, 1999a, 1999b, 2000a, 2000b) to describe the dynamic and radiation process of the afterglows of
GRB 081029. These equations are applicable to both radiative and adiabatic blastwaves, and are
appropriate for both ultra-relativistic and non-relativistic stages (Huang et al. 1999a, 1999b, 2000a, 2000b, 2003).
Most importantly, it takes the lateral expansion, the cooling of electrons, and the equal arrival time
surface effect into consideration. The evolution of radius ($R$), the swept-up mass ($m$), the half
opening angle $(\theta)$ and the Lorentz factor are described as:
\begin{equation}
\frac{dR}{dt}=\beta c \gamma(\gamma+\sqrt{\gamma^{2} - 1}),
\end{equation}
\begin{equation}
\frac{dm}{dR}=2 \pi R^{2}(1-\cos\theta) n m_{p},
\end{equation}
\begin{equation}
\frac{d\theta}{dt}=\frac{c_{s}(\gamma+\sqrt{\gamma^{2} - 1})}{R},
\end{equation}
\begin{equation}
\frac{d\gamma}{dm}=\frac{-(\gamma^{2}-1)}{M_{ej}+\epsilon
m+2(1-\epsilon)\gamma m},
\end{equation}
where $\beta=\sqrt{\gamma^{2} - 1}/\gamma$, $n$ is the number density of surrounding ISM,
$m_{p}$ is the mass of the proton, $c_{s}$ is the co-moving sound speed, $R$ is the distance
from the center in the burster's frame, $t$ is the observer's time, $M_{ej}$ is the initial ejecta mass,
$m$ is the swept-up ISM mass, and $\epsilon$ is the radiative efficiency.
A reasonable expression for $c_{s}$ is:
\begin{equation}
c_{s}^{2}=\hat{\gamma}(\hat{\gamma}-1)(\gamma-1)\frac{1}{1+\hat{\gamma}(\gamma-1)}c^{2},
\end{equation}
where $\hat{\gamma}\approx(4 \gamma+1)/(3 \gamma)$ is the adiabatic index
(Dai et al. 1999).

In the standard fireball model, as the blast wave sweeps up the surrounding medium, the shock
accelerates electrons. The afterglow emission arises from synchrotron radiation of these
shocked electrons due to their interaction with magnetic field. Considering the energy injection,
the differential equations should be modified accordingly so that it can be applicable to our case.
Due to strong magnetic field and rapid rotation, a new-born millisecond pulsar will lose its
rotation energy through magnetic dipole radiation. Dai \& Lu (1998) have considered the energy
injection from a new-born strongly magnetized millisecond pulsar at the center of GRB. They argued
that the radiation power evolves with time as $L(t) = L_{0}(1+t/T)^{-2}$, where $L_{0}$ is the initial
luminosity, $t$ is the time in the burster's rest frame, and $T$ is the spin-down timescale.
Considering an adiabatic relativistic hot shell which receives the energy injection from the
central engine through a Poynting-flux-dominated flow, Zhang \& ${\rm M\acute{e}se\acute{a}ros}$
(2001b) gave an intrinsic luminosity law, $e.g.$ $L(t) \propto t^{q}$, where $t$ is the intrinsic
time of the central engine. They pointed out that usually $q = 0$ during the injection phase in
many cases. To explain the special behaviors of GRB 070610 in the observed X-ray and optical afterglow
light curves, Kong \& Huang (2010) assumed that the energy injection power takes the form of
$dE_{inj}/dt = Q t^{q}$ for $t_{start}<t<t_{end}$, where Q and q are constants, $t_{start}$ is
the beginning time of the energy injection, and $t_{end}$ is the ending time of the energy injection.
For some types of central engines, such as a black hole plus a long-lived debris torus system,
the energy injection to the fireball may in principle continue for a time scale significantly
longer than that of the gamma-ray emission (Zhang \& ${\rm M\acute{e}se\acute{a}ros}$ 2001a).
Taking into account all the energy injection forms as described above, and the extremely fast
optical rebrightening of the afterglow of GRB 081029 at about 3000 s after the trigger time,
here we take the same form of energy injection power as Kong \& Huang (2010). The differential
equation for the evolution of the bulk Lorentz factor (i.e. Eq. (4) ) should then be changed to:
\begin{equation}
\frac{d\gamma}{dt}=\frac{1}{M_{ej}+\epsilon m+2(1-\epsilon) \gamma
m}\times(\frac{1}{c^{2}} \frac{dE_{inj}}{dt}-(\gamma^{2}-1) \frac{dm}{dt}).
\end{equation}
In the simplest case, q = 0.

\section{NUMERICAL RESULTS}

Paying special attention to the rebrightening of the afterglow of GRB 081029 in the optical band
at around 3000 s, we use the energy injection model to calculate the X-ray and optical afterglow
light curves in detail, and compare the numerical results with the observations.
As the fluence of GRB 081029 measured in the 15 --- 350 ${\rm keV}$ energy range by BAT is
$2.1 \pm 0.2 \times 10^{-6}$ ${\rm erg~cm^{-2}}$ (Nardini et al. 2011), the isotropic energy
released in the rest-frame in the 15 --- 350 ${\rm keV}$ band is then $E_{0,iso}$ = $3.1\times10^{53}$
${\rm erg}$. In our calculations, we will assume this value as the initial isotropic kinetic 
energy of the outflow. Other parameters are taking as following: the initial Lorentz
factor of the blast wave $\gamma_{0}$ = 70, the ISM number density $n$ = 2.0 ${\rm cm^{-3}}$,
the power-law index of the energy distribution of electrons $p=2.4$, the luminosity distance of
the source $D_{L}$ = $3.5\times10^{7}$ ${\rm kpc}$, the electron energy fraction
$\epsilon_{e}$ = 0.04, the magnetic energy fraction $\epsilon_{B}$ = 0.004, the initial half
opening angle of the ejecta $\theta=0.04$ radian, and the observing angle $\theta_{obs}$ = 0,
where the observing angle is defined as the angle between the line of sight and the jet axis.

For the significant rebrightening at approximately 3000s after the trigger, we assume an energy
luminosity with $Q$ = $7.0\times10^{47}$ ${\rm erg}$ ${\rm s^{-1}}$, $q$ = 0, $t_{start}$ =
$2.8\times10^{3}$ s, and $t_{end}$ = $5.3 \times10^{3}$ s. This will lead to a total energy
injection of $E_{inj}$ = 3.1 $E_{0}$, where $E_{0}$ = $(1-\cos\theta)E_{0,iso}$
is the collimation-corrected energy. According to the analysises by Zhang
\& ${\rm M\acute{e}se\acute{a}ros}$ (2002), such an injected energy higher than that of the 
original kinetic energy of the outflow should be able to generate an obvious rebrightening 
in the afterglow lightcurve. To get the best fit to the observations
of GRB 081029 in the optical band, another energy injection process is required, which
gives birth to the observed flat stage occurring between about 8000s and 13000s.  The parameters
corresponding to this second energy injection are: $Q$ = $3.5\times10^{47}$ ${\rm erg}$ ${\rm s^{-1}}$,
$q$ = 0, $t_{start}$ = $8.0\times10^{3}$ s, and $t_{end}$ = $1.3\times10^{4}$ s.
Additionally, contribution from a host galaxy with the magnitude of $r^{\prime}\sim25$ ${\rm mag}$
is assumed, which will account for the final flat stage of the optical afterglow.

Using the model and parameters described above, we can give a satisfactory fit to the observed X-ray
and optical afterglow of GRB 081029. Figure 1 illustrates the observed optical data of GRB 081029,
taken from Nardini et al (2011). Also plotted are our calculated $R$ band flux densities ($S_{R}$).
We see that the observed optical afterglow light curve can be satisfactorily reproduced.


\begin{figure}
   \vspace{0.5cm}
   \begin{center}
   \plotone{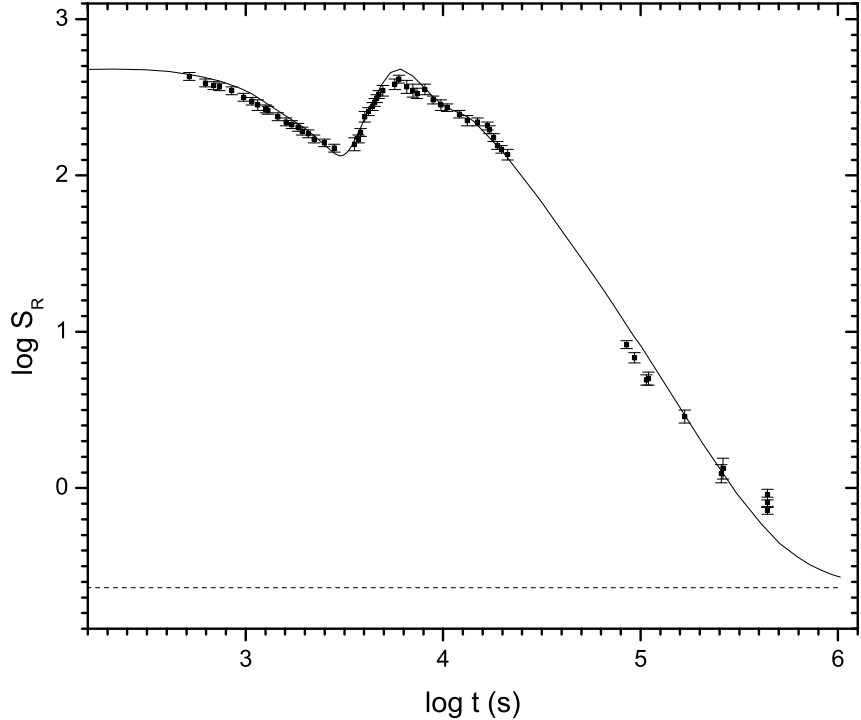}
   \caption{Numerical fit to the optical afterglow (in units of ${\rm \mu Jy}$)
   of GRB 081029 by using the two-step energy injection model. The observational data are
   taken from Nardini et al (2011). The solid line is our theoretical optical afterglow
   light curve corrected for extinction. The dashed line is the contribution from a
   host galaxy with the magnitude of 25 mag.}
   \end{center}
   \label{Fig:plot1}
   \end{figure}

Figure 2 illustrates the observed X-ray light curve ($F_{X}$) of GRB 081029 in the 0.3 --- 10 ${\rm keV}$ band.
Observational data are taken from Nardini et al (2011).
No rebrightening as significant as in the optical band could be identified. But Holland et al. (2011)
argued that there could be some flares in the observed X-ray light curve between approximately 2500 s and
5000 s, and the time scale of the flares were  $\Delta t/t < 1$. However, the error bars of the observational
data are generally large (as compared with optical data), so that no firm conclusion could be drawn.
Also plotted in Figure~2 is our theoretical light curve by using the same energy injection model with
the same parameters as in Figure 1. It is interesting to note that in X-ray band, the theoretical amplitude
of the rebrightening due to energy injection is much smaller as compared with that at optical wavelength.
Our numerical results are then actually well consistent with the observed X-ray light curve.

It should be noted that in our calculations, we do not consider the reverse shock emission component 
during the energy injection stage. Actually, depending on different types of central engines,
the injected energy could be in a kinetic-energy-dominated form or in a Poynting-flux-dominated form (Usov 1994; 
${\rm M\acute{e}se\acute{a}ros}$ \& Rees 1997b). When the injected energy is of a kinetic form, not of a Poynting 
flux, then during the injection process, reverse shocks might be formed . Emission from such reverse shocks 
could significantly enhance the rebrightenings (Zhang \& ${\rm M\acute{e}se\acute{a}ros}$ 2002). 
In realistic cases, it is also possible that when the fast shell catches up with the slow shell and gives 
birth to an energy injection, the relative speed between the two colliding shells is not too high, so that 
only a mildly relativistic reverse shock is generated (Rees \& ${\rm M\acute{e}se\acute{a}ros}$ 1998;
Kumar \& Piran 2000; (Zhang \& ${\rm M\acute{e}se\acute{a}ros}$ 2002)). In this case, emission from the 
reverse shock will not be very strong. In our modeling, both of the two energy injections last for thousands 
of seconds. Although the injection flows are assumed to be kinetic-energy-dominated, we believe that 
the collisions during the injection process would not be too violent and the induced reverse shock 
would not be too strong. So we have omitted the emission component of the reverse shock. In the future, 
when more detailed studies are carried out, the effects of the reverse shock should be included.

\begin{figure}
   \vspace{0.5cm}
   \begin{center}
   \plotone{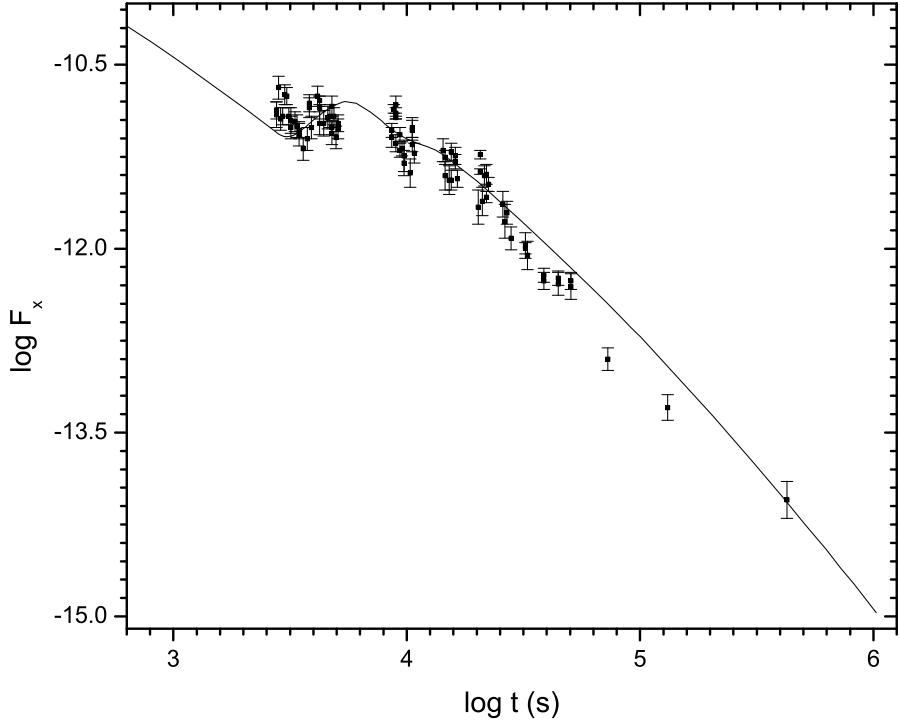}
   \caption{Our numerical fit to the X-ray afterglow (in units of ${\rm erg~cm^{-2}
   ~s^{-1}}$) of GRB 081029 by using the two-step energy injection model. The
   observed data points are taken from  Nardini et al (2011). The solid line is our theoretical
   light curve for GRB 081029 by using the same model as in Figure~1.}
   \end{center}
   \label{Fig:plot2}
   \end{figure}

\section{CONCLUSION AND DISCUSSION}
\label{sect:discussion}

GRBs are widely believed to be produced by relativistically expanding blastwaves at
cosmological distances. It is possible to observe early afterglows of many GRBs in the
first few hours after the trigger due to the launch of the Swift satellite. Many remarkable
and unexpected features such as rebrightenings in the optical afterglows have been found,
challenging the view that the optical afterglow light curves should be smooth
(Laursen \& Stanek 2003), which is formed since the discovery of
the first gamma-ray burst afterglow (Sahu et al. 1997).
GRB 081029 is characterized by a complex optical light curve. A distinguishing feature of
this event is the obvious rebrightening in the optical band at around 3000 s after the burst.
In this paper, we calculate the overall dynamical evolution of the blastwave numerically by
adopting the energy injection model to reproduce the X-ray and optical afterglow light curves
of GRB 081029. We assume that the relativistic shock expands in a uniform ISM. We show that
the remarkable rebrightening observed in optical band can be satisfactorily modeled by our model.
We argue that the rapid rise is due to the energy injection from the late-time interaction of a
slow shell with the forward shock. In fact, similar mechanism of energy injection has also
been used to explain the afterglows of some other GRBs, such as GRB 010222
(${\rm Bj\ddot{o}rnsson}$ et al. 2002), GRB 021004 (${\rm Bj\ddot{o}rnsson}$ et al. 2004),
GRB 021004 (de Ugarte Postigo et al. 2005), GRB 030329 (Huang et al. 2006) and GRB 051221A
(Fan \& Xu 2006). In our calculations, many of the parameters, such as the power-law index of
the energy distribution of electrons ($p$), the electron energy fraction ($\epsilon_{e}$),
the magnetic energy fraction ($\epsilon_{B}$), the initial half opening angle of the ejecta
($\theta$), have been evaluated typically.

In our model, we have assumed two periods of energy injection, each with a constant injection power.
One injection starts at $2.8\times10^{3}$ s and ends at $t_{end}$ = $5.3 \times10^{3}$ s, with a
power of $Q$ = $7.0\times10^{47}$ ${\rm erg}$ ${\rm s^{-1}}$. This energy injection is mainly engaged to
account for the rapid rebrightening at about 3000 s. The other injection starts at
$8.0\times10^{3}$ s and ends at $t_{end}$ = $1.3\times10^{4}$ s, with the power being
$Q$ = $3.5\times10^{47}$ ${\rm erg}$ ${\rm s^{-1}}$. This energy injection can help to explain
the slight rebrightening at about 10000 s. Physically, this kind of energy injections can
be produced by the fallback of materials onto the central compact object of the burster. The fallback
is usually continuous, but clumps sometimes could exist in the falling material. When a large clump
suddenly plunges into the accretion disk, the accretion rate can be significantly increased, giving
birth to a strong outflow. The relativistic shell resulted from the energetic outflow moves outward
at approximately a constant speed in a dilute environment that has been swept-up by the previous
external shock. It can finally catch up with the fireball material and inject the energy into
the fireball, producing a significant rebrightening in the afterglow.

In our fitting to the optical afterglow of GRB 081029, extinction has been taken into account.
The theoretical light curve of GRB081029 in the optical band was shifted downward by about 1.57 mag.
It is consistent with the result derived by Holland et al. (2012) who suggested that the rest
frame $V$ band extinction is $A_{V}$ $\leq$ 2 mag. Extinction has also been considered in many other GRBs.
Sokolov et al (2001) pointed out that there is a significant internal extinction in the host galaxies
of GRB 970508, GRB 980613, GRB 980703, GRB 990123 and GRB 991208. Rol et al (2007)
suggested a high internal extinction, at least 2.3 magnitudes at the infrared (J) wavelength and 5.4
magnitudes at U Band in the rest-frame to explain the absence of an optical afterglow for
GRB 051022, which is a prototypical dark burst. For high redshift GRBs, Draine (2000) draw the conclusion
that absorption by vibrationally-excited $H_{2}$ could be responsible for the pronounced drop in flux
between R and I band. Considering the fact that the redshift of GRB 081029 is about z = 3.8479,
absorption by $H_{2}$ may be another reason for the phenomenon that the observed optical flux density
of GRB 081029 is much less than the theoretical value. Kong et al (2009) derived
the extinction of the host galaxy of GRB 980703 as $A_{V}\sim2.5$ ${\rm mag}$ by modeling
the multi-band afterglow light-curves.

In conclusion, we have shown that our model can reasonably explain both the X-ray and optical
afterglow light curves of GRB 081029. Especially the observed optical rebrightening can be
fitted quite well by assuming a constant energy injection. In the future, more detailed studies
on the energy injection processes will be helpful to provide important clues on the origin and
the trigger mechanism of GRBs.

\begin{acknowledgements}
This work was supported by the National Natural Science Foundation of China
(Grant No. 11033002) and the National Basic Research Program of China (973 Program, Grant No. 2009CB824800).
\end{acknowledgements}

\label{lastpage}

\end{document}